# Simulation-Driven Design of a Multilayer Plasmonic Sensor Using Cu-Ni and BaTiO$_3$ for Waterborne Pathogen Detection

R. Runthala, V. K. Venkatesh, D. Gupta, and P. Arora, *Senior Member, IEEE*

*Abstract*— We present a simulation-guided design for a multilayer surface plasmon resonance (SPR)-based biosensor that can detect changes in the refractive index of a target induced by analytes. Surface plasmons are excited through a hybrid Kretschmann configuration with a low-refractive-index calcium fluoride (CaF$_2$) prism under transverse magnetic polarization illumination. In the sensing architecture, the copper (Cu) layer is a plasmonic metal, overlaid with a thin nickel (Ni) layer that protects it from oxidation. To improve analyte coupling and electromagnetic confinement, a dielectric layer of barium titanium oxide (BaTiO$_3$) and a monolayer of graphene oxide (GO) are used. The layer structure is iteratively optimized employing the transfer matrix method for angular interrogation at a wavelength of 1064 nm, focusing on key performance metrics including sensitivity, minimum reflectivity, and figure of merit (FOM). The finite element method-based simulation confirms SP excitation through consistent and optimal performance of individual layer thicknesses with Cu thickness of 30 nm and BaTiO$_3$ thickness of 5 nm. The SPR-based sensor (CaF$_2$-Cu-Ni-BTO-GO) is designed and has achieved a sensitivity of 157.8°/RIU and an FOM of 17.48 RIU$^{-1}$ while detecting the presence of the E. Coli bacterium in the water, indicative of the biosensor's application.

## I. Introduction

The ongoing advancement of biosensor technologies has significantly transformed the monitoring of biological systems, allowing for label-free and highly sensitive detection in clinical diagnostics, environmental monitoring, and biotechnological applications. Contemporary sensing platforms are now incorporated into diverse fields, including satellite-based remote sensing systems, wearable health monitors, and intracellular tools like cellular rovers [1-4]. This swift expansion into various application areas has resulted in the development of sensing systems that are not only compact and cost-effective but also capable of providing high precision in complex and ever-changing biological environments [5,6].

Among all biosensing mechanisms, optical sensors based on surface plasmon resonance (SPR) have received the most interest, with SPR being the most sensitive in relation to a change in refractive index at a metal-dielectric interface [7-9]. SPR occurs when a p-polarized light excites the surface plasmons, which are the collective oscillations of the conduction electrons, at the interface between a thin metal film and a dielectric medium. As a result of this interaction, there is a pronounced dip in the reflectance at the resonance angle [10]. The position and intensity of this dip are extremely sensitive to a small change in the refractive index at the surface. This remarkable sensitivity of the SPR sensor makes it well suited for applications in biosensing, medical diagnostics, and chemical detection [11].

The Kretschmann configuration continues to be the most commonly used setup for SPR sensing. In this arrangement, a high-refractive-index prism is used to excite surface plasmons on a thin metallic film coated on its base. Noble metals like gold (Au) and silver (Ag) have traditionally been favored for this purpose because of their excellent optical characteristics [12]. However, their high cost and the difficulty of scaling them in an economically viable way have motivated the exploration of alternative plasmonic materials [13]. As a result, research efforts have turned toward more abundant and affordable metals such as aluminum (Al) and copper (Cu), which offer a practical trade-off between performance and economic feasibility [14,15]. Among the available alternatives, Cu stands out as a particularly attractive candidate because of its excellent electrical conductivity, widespread availability, and robust plasmonic response in the visible and near-infrared parts of the spectrum [13].

Many studies show that Cu-based plasmonic devices can rival, and sometimes even surpass, those based on Au and Ag, especially when the structural parameters are meticulously optimized. In specific configurations, Cu films have been shown to support enhanced confinement of electromagnetic fields and to produce sharper resonance features [15]. Muthumanikkam et al. demonstrated that integrating a suitable dielectric layer with a Cu film can substantially improve sensitivity [15]. The plasmonic response of Cu can be enhanced by integrating two-dimensional (2D) materials like graphene or graphene oxide (GO). These materials enhance the interaction between light and matter through π–π stacking and also create extra optical absorption pathways, resulting in better overall sensor performance [16].

While Cu offers several advantages, a key limitation in SPR-based applications is its susceptibility to oxidation under ambient conditions. This oxidative degradation greatly affects the sensor's stability over time as well as its plasmonic performance. Researchers have suggested different protective methods to maintain Cu's optical properties over time in response to this issue. To minimize oxidation, researchers have explored the utilization of another metal layer, a dielectric film, or 2D nano materials over Cu [17]. These layer coatings have effectively reduced oxidation without affecting the electromagnetic coupling at the metal–dielectric interface [16].

P. Arora is with the Department of Electrical & Electronics Engineering, Birla Institute of Technology and Science, Pilani, Rajasthan 333031 India (Corresponding author: +91 7976790062; e-mail: pankaj.arora@pilani.bits-pilani.ac.in).

V. K. Venkatesh, D. Gupta and R. Runthala are with Birla Institute of Technology and Science, Pilani, Rajasthan 333031 India (e-mail: f20220462@pilani.bits-pilani.ac.in, f20221206@pilani.bits-pilani.ac.in, and p20230044@pilani.bits-pilani.ac.in).

To further improve sensor performance, multilayered structures with high-permittivity dielectrics and engineered nanostructures, including periodic hole arrays and deep nanogroove patterns, have been developed [18,19]. These geometrical modifications amplify the local electric field and increase the effective surface area available as binding sites for the analyte. As a result, both the sensitivity and the quality factor of the SPR response are significantly improved [20-22]. Concurrently, graphene oxide (GO), an atomically thin and oxygen-functionalized form of graphite, has attracted significant interest in optical sensing due to its large specific surface area and adjustable electrical conductivity that responds to surface adsorption. These characteristics enhance its efficacy in identifying a diverse range of molecular targets, encompassing both gaseous substances and intricate biomolecules [23].

Together with advances in experimental research, computational modeling has become increasingly important in the design of SPR sensors. Techniques such as the transfer matrix method (TMM) and finite element method (FEM) have proven to be effective tools, enabling detailed simulation of reflectance spectra, electromagnetic field distributions, and mode coupling behavior. By adjusting material properties and structural configurations within these models, researchers can gain deep insights into how each parameter shapes sensor performance, leading to more refined and effective designs [24]. Optimization studies frequently involve adjusting layer thicknesses, tuning incident wavelengths, and varying analyte refractive indices to identify designs that deliver sharp resonance dips, narrow full-width-at-half-maximum (FWHM) values, and elevated figures of merit (FOM) [25].

This study proposes a simulation-guided design of a multilayer SPR biosensor that includes a Cu film coated with a thin nickel layer, followed by a BTO dielectric layer and a functional top coating of graphene oxide. The structure consists of a calcium fluoride ($CaF_2$) prism in a modified Kretschmann configuration and is designed to operate at a near-infrared wavelength of 1064 nm. Angular reflectance characteristics are calculated using the TMM in MATLAB, while FEM simulations in COMSOL Multiphysics are used to confirm the presence of SPR Mode. The sensor architecture is optimized by varying individual layer thicknesses and the number of GO monolayers, using parameters such as minimum reflectance and FOM to guide the refinement of the design. Post optimization, the final sensor configuration is calibrated to detect Escherichia coli (E. coli) using SPR as the transduction mechanism.

## II. SENSOR DESIGN AND OPTIMIZATION

The proposed SPR sensor is constructed as a multilayer heterostructure based on the Kretschmann configuration. A calcium fluoride ($CaF_2$) prism with a refractive index of 1.426 is chosen for efficient coupling of p-polarized incident light due to its chemical inertness [25]. The sensor operates at a wavelength of 1064 nm, which offers a trade-off between the visible and near-infrared regions. This choice allows for effective field penetration into the analyte and guarantees compatibility with standard optoelectronic devices. Figure 1(a) shows the schematic for the proposed multilayer plasmonic sensor.

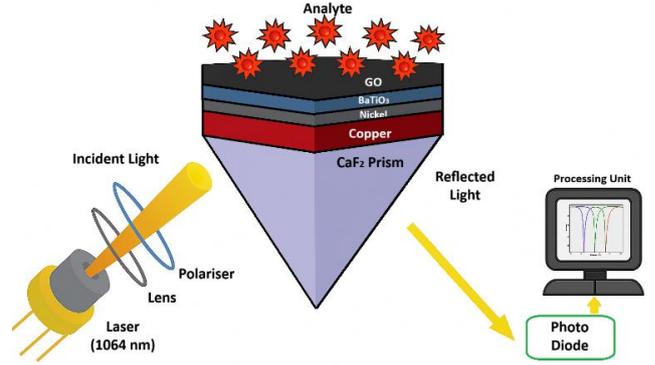

Figure 1. Schematic (not to scale) of the proposed four-layer plasmonic sensor using the Kretschmann configuration.

The sensor is engineered to evaluate its optical performance using three main metrics: sensitivity, minimum reflectance, and FOM. Sensitivity is the ratio of the change in resonance angle to the change in the analyte's refractive index. The FWHM refers to the width of the resonance curve measured at 50% of the reflected intensity and provides insight into the sharpness of the resonance. The FOM is calculated as the ratio of sensitivity to FWHM. These parameters collectively provide a clear picture of how well the sensor can detect changes in the surrounding medium. These parameters guide the optimization process, which is carried out using TMM under the angle interrogation scheme. Water is used as the analyte, and its refractive index varies from 1.33 to 1.34 to simulate realistic sensing conditions.

Multiple simulations were performed using different combinations of Cu and BTO layer thicknesses to identify the optimal thickness of the Cu and BTO layers, as shown in Table I. A 5 nm Ni layer is also considered over the Cu surface to prevent Cu from oxidation during the Cu-BTO optimization. FOM and minimum reflectance values improve as the Cu thickness increases to 25 nm. Beyond this point, the reflectance dip starts to rise. A Cu thickness of 30 nm is selected as the optimal value, providing the highest FOM while keeping minimum reflectance below 0.05, which ensures a balanced trade-off between high performance and signal quality. The thickness of the BTO layer is set to 5 nm. Higher thicknesses were found to weaken field localization in the analyte, which is suboptimal.

TABLE I. SENSING PARAMETERS WITH DIFFERENT VALUES OF CU AND BTO LAYER THICKNESS FOR OPTIMIZATION

| S.N. | Structure | FOM [$RIU^{-1}$] | $R_{min}$ [a.u.] |
|---|---|---|---|
| 1 | BTO=5nm, Cu=20nm | 12.62 | 0.0400 |
| 2 | BTO=5nm, Cu=25nm | 14.51 | 0.0000 |
| 3 | BTO=5nm, Cu=30nm | 17.69 | 0.0395 |
| 4 | BTO=5nm, Cu=35nm | 23.93 | 0.1454 |
| 5 | BTO=5nm, Cu=40nm | 37.39 | 0.2875 |
| 6 | BTO=10nm, Cu=20nm | 13.12 | 0.0479 |

| 7 | BTO=10nm, Cu=25nm | 14.85 | 0.0200 |
| 8 | BTO=10nm, Cu=30nm | 17.62 | 0.0351 |
| 9 | BTO=10nm, Cu=35nm | 22.47 | 0.1391 |
| 10 | BTO=10nm, Cu=40nm | 32.52 | 0.2811 |
| 11 | BTO=15nm, Cu=20nm | 13.83 | 0.0544 |
| 12 | BTO=15nm, Cu=25nm | 15.73 | 0.0700 |
| 13 | BTO=15nm, Cu=30nm | 18.41 | 0.0324 |
| 14 | BTO=15nm, Cu=35nm | 22.65 | 0.1361 |
| 15 | BTO=15nm, Cu=40nm | 30.71 | 0.2792 |

For the graphene oxide (GO) layer, the number of layers varies from 1 to 10, each 1nm thick, while keeping the thicknesses of the other layers fixed. Figure 2 shows that increasing the number of GO layers results in broader FWHM and higher minimum reflectance, both of which reduce sensing performance.

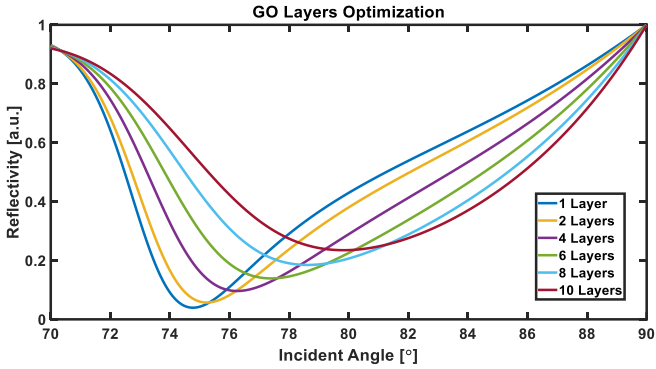

Figure 2. Effect of GO layers on the sensor's performance

Thus, a single GO monolayer (1 nm) is selected as the optimal configuration. The final sensor structure is summarised in Table II with the refractive index taken from [24, 26-29].

TABLE II. REFRACTIVE INDEX VALUES OF MATERIAL USED IN MULTILAYER ENGINEERED DEVICE WITH OPTIMIZED THICKNESS

| S.N. | Material | Refractive index [RIU] | Thickness [nm] |
|---|---|---|---|
| 1 | CaF$_2$ Prism | 1.426 | – |
| 2 | Cu | 0.2689 - 6.9575i | 30 |
| 3 | Ni | 2.8955 - 5.2413i | 5 |
| 4 | BTO | 2.254 | 5 |
| 5 | GO | 2.895 - 1.7993i | 1 |

To validate the electromagnetic field behavior of the optimized structure, simulations were performed using the Finite Element Method (FEM) in COMSOL Multiphysics at a resonance angle of 74°. The electric field intensity was observed to be highest at the metal–dielectric interface, indicating strong surface plasmon excitation and effective field penetration into the analyte region. The sensor model was constructed by assigning appropriate materials to each layer, and TM-polarized light was introduced from the prism side. Scattering boundary conditions were applied to all exterior faces to minimize reflection effects. The electromagnetic waves with a frequency domain module were used to carry out a frequency domain study following mesh generation. The resulting electric field distributions were analyzed to examine the resonance behavior and confirm the efficiency of field confinement at the sensing interface.

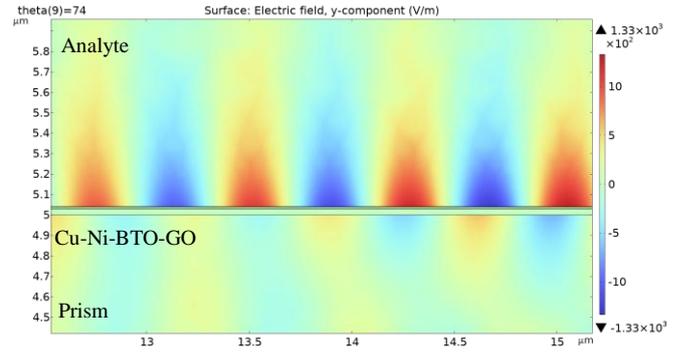

Figure 3. Electric field distribution at the SPR angle for the proposed plasmonic sensor

### III. ENGINEERED SPR SENSOR FOR E. COLI DETECTION

To assess the sensor's biosensing capability, the refractive index of the analyte solution is varied to simulate increasing concentrations of β-D-galactosidase, the biomarker secreted by *E. coli*. The refractive index of the analyte solution was modelled using a linear relationship between the refractive indices of water and β-D-galactosidase, scaled by enzyme concentration. In this approach, water and β-D-galactosidase are assigned refractive indices of 1.326040 and 1.36294, respectively [30], and the enzyme concentration is varied from 0% to 4%. This model enables direct estimation of the effective refractive index at each concentration level and allows for the simulation of the sensor's optical response under biologically relevant conditions. The resulting variation is shown in Figure 4, which illustrates the linear dependence of refractive index on enzyme concentration across the examined range.

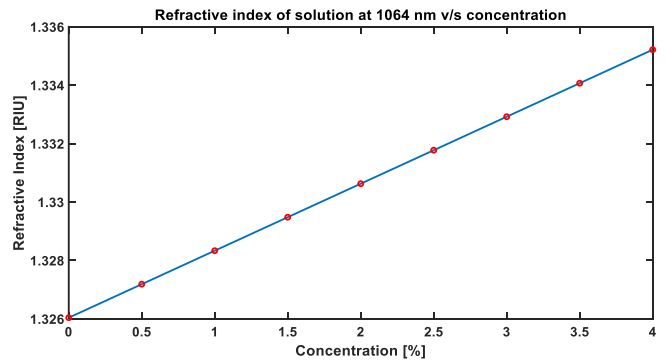

Figure 4. Variation of refractive index with changing concentration of β-D-galactosidase at 1064nm

The sensor's reflectance response to each concentration is simulated using TMM. The angular reflectance curves are computed for each concentration, and the resonance angle shift is observed. As shown in Figure 5, the resonance dip exhibits a redshift with increasing enzyme concentration, confirming the sensor's sensitivity to refractive index changes in the analyte environment.

From the reflectance curves, the resonance angle, sensitivity, and FOM are extracted and summarized in Table III. The sensor demonstrates a high angular sensitivity of around 157.8 °/RIU while maintaining a FOM of approximately 17.48 RIU$^{-1}$. These results highlight the sensor's capability for detecting subtle biomolecular changes in aqueous media.

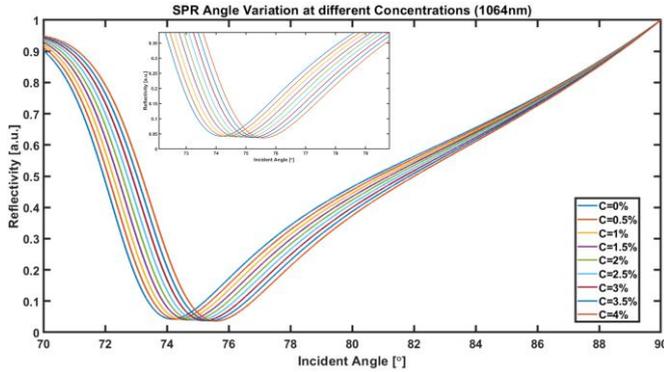

Figure 5. Reflectance characteristics of the engineered sensor for Varying Concentrations of β-D-Galactosidase

TABLE III. FINAL SPR SENSOR CHARACTERISTICS EXTRACTED FROM REFLECTANCE CURVES

| Concentration [%] | Refractive index (RIU) | $\theta_{min}$ [°] | Sensitivity (°/RIU) | FOM (RIU$^{-1}$) |
|---|---|---|---|---|
| 0.0 | 1.326040 | 74.175 | - | – |
| 0.5 | 1.327187 | 74.347 | 150.00 | 16.72 |
| 1.0 | 1.328334 | 74.519 | 150.00 | 16.69 |
| 1.5 | 1.329481 | 74.693 | 151.70 | 16.86 |
| 2.0 | 1.330628 | 74.868 | 152.57 | 16.94 |
| 2.5 | 1.331775 | 75.045 | 154.38 | 17.13 |
| 3.0 | 1.332922 | 75.224 | 156.06 | 17.30 |
| 3.5 | 1.334069 | 75.404 | 156.93 | 17.39 |
| 4.0 | 1.335216 | 75.585 | 157.80 | 17.48 |

A consistent and nearly linear increase in the SPR angle with increasing enzyme concentration is observed, as shown in Table III and Figure 6. This trend is particularly important in biosensing applications, as it indicates that the sensor's optical response scales predictably with analyte concentration.

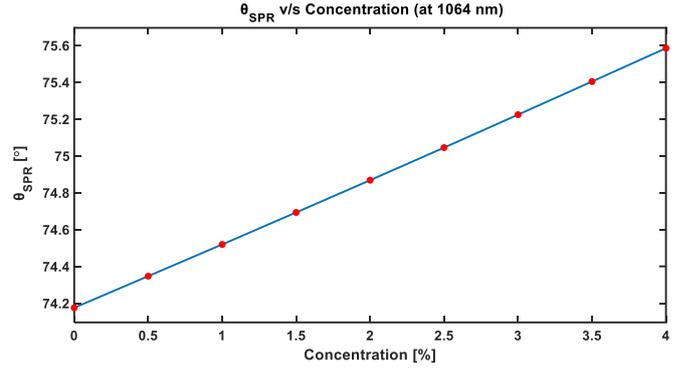

Figure 6. Variation of SPR angle with changing concentration of β-D-galactosidase at 1064nm

These findings confirm that the proposed SPR sensor can serve as a dependable and sensitive platform for detecting *E. coli* indirectly via β-D-galactosidase concentration tracking.

## IV. CONCLUSION

This study presents a multilayer SPR sensor that has been engineered for refractive index-based biosensing using a simulation-based approach. The final CaF$_2$–Cu–Ni–BTO–GO configuration, operating at 1064 nm, achieved strong optical performance, with a sensitivity of approximately 154°/RIU and a FOM of 17.5 RIU$^{-1}$. Calibration done using β-D-galactosidase demonstrated the sensor's capability for indirect E. coli detection in aqueous media, highlighting its potential for label-free pathogen sensing. Future studies could include surface functionalization to improve specificity toward E. coli, using materials such as organic polymers, nanowires, or two-dimensional nanomaterials. Techniques like surface-enhanced infrared absorption to enhance molecular binding and interactions at the sensor surface could also be explored.

ACKNOWLEDGMENT

P. A. would like to acknowledge DBT (BT/PR48652/MED/32/861/2023) and BITS Pilani (CDRF and SPARKLE) for the financial support.